\documentclass[12pt]{article}
\usepackage{latexsym,amsmath,amssymb,amsbsy,graphicx}
\usepackage[space]{cite} 

\usepackage{epsfig}

\makeatletter\def\@biblabel#1{\hfill#1.}\makeatother
\allowdisplaybreaks \multlinegap0pt 

\begin{document}

\begin{center}

{\Large{Critical Charge in Gapped Graphene: the role of the screening of the interaction potential by $\sigma$-orbitals.}}\\[9pt]

{\large A. A. Novoselov $^{1,2 \, a}$, O. V. Pavlovsky $^{1,2,3 \, b}$}\\[6pt]

\parbox{.96\textwidth}{\centering\small\it
$^1$ Faculty of Physics, Moscow State University, Moscow, Russia \\
$^2$ Institute for Theoretical and Experimental Physics, Moscow, Russia\\
$^3$ National Research Nuclear University "MEPhI", Moscow, Russia\\

E-mail: $^a$novoselov@goa.bog.msu.ru $^b$ovp@goa.bog.msu.ru}\\[1cc]
\end{center}

{\parindent5mm Due to its unique structure, graphene provides a
condensed-matter model of particle physics phenomena. One is the
critical charge which is highly interested. The investigation of
critical charge in gapped graphene is performed within  single
particle approach by means of Dirac equation integration. The
screened Coulomb interaction between charged defect and graphene
electron excitations is investigated. Two kinds of mass gap
generation and various values of substrate dielectric
permittivities are considered. It is shown that the critical
charge phenomenon can be observed  even with quite small charges
for physically motivated parameters. \vspace{2pt}\par}




\section*{Introduction}

The   critical charge problem is significant for modern physics.
In nuclear physics this phenomenon is responsible for the
instability of nuclei with charges large than some critical values
\cite{Pom}, \cite{Zel}. The fundamental reason of it is the
relativistic behavior of electrons in the strong potential of a
heavy nucleus. The idea of the experimental investigation of
supercharges in heavy ion collision experiments was proposed many
years ago \cite{Popov1}, \cite{Popov2}, \cite{Raf}.  But, after
great efforts,  the situation with the experimental observation of
critical charge phenomenon in nuclear physics is still unclear.
The first observation of the critical charge evidence was reported
in \cite{exp1}, \cite{exp2}, but later experiments did not confirm
the first results \cite{EPOS}, \cite{EPOS2}, \cite{APEXa},
\cite{APEXb}, \cite{APEXc},\cite{APEXd}. The modern theoretical
studies \cite{Greiner} show that electron-positron pier creation
in heavy ion collisions is a more complicated process than it was
expected earlier, and the possibility of the critical charge
observation is still an open problem.

The phenomenon of critical charge can be considered not only in
nuclear physics. The one-dimensional Coulomb problem in carbon
nanotubes was considered in \cite{tube}. The critical charge has
recently become possible to be explored also in two-dimensional
condensed matter structures, e. g. in graphene
\cite{Wang},\cite{Nature}.

  Graphene is a single layer of carbon atoms \cite{Katsnelson}. One of its
most significant features is that the electron excitations are
massless fermions and can be described as effective
quasi-relativistic Dirac particles. The "speed of light" is
determined by the Fermi speed $v_f$ - the speed of elementary
excitation propagating, and it turns to be much smaller than the
 light speed in vacuum. This fact makes the studies of the
quasi-relativistic phenomena in graphene much easier.
Particularly, the critical charge in graphene is expected to be
obtained at smaller values of $Z$ \cite{Shitov1}, \cite{Pereira}.
After these theoretical considerations the critical charge
phenomenon has been observed experimentally \cite{Wang}. The
quasi-bound states of itinerant graphene electrons at
supercritically charged vacancies were recently detected in
graphene on the boron nitride (BN) substrate \cite{Nature}. The
interesting feature of this experiment stems from the fact that
mass gap is generated by BN substrate. In our studies we consider
a similar system.

 The great progress in the study of critical charge problem
in graphene can help us in the study of the critical charge
problem in nuclear physics. Unfortunately, the analogy is not full
due to a very essential  difference: the excitations in graphene
are massless particles but electrons in QED have non-zero masses.
If we want to use graphene as a condensed matter model of the
processes in strong electromagnetic fields in atomic physics, some
mechanism of mass gap generation should be proposed.

The obtaining of gapped graphene is a very considerable problem
for future nanotechnological applications. The mass gap in
graphene can be induced by the sublattice symmetry breaking of the
hexagonal lattice. Such a breaking can be achieved by the
interaction with special substrates if one of the sublattices
interacts with the substrate stronger than the another one. For
example, the hexagonal boron nitride has almost the same lattice
as graphene. If one places a graphene layer on the boron nitride
substrate, its sublattices will be not equivalent to each other
due to the difference between  the interactions of carbon atoms
with atoms of boron and atoms of nitrogen. This sublattice
symmetry breaking produces the mass gap about 53 meV \cite{BN}.
Another prospective possibility to produce the gapped graphene was
proposed in \cite{SiC}. In this case substrate is SiC but graphene
layers are epitaxially grown on this SiC substrate. It was shown
that the mass gap about 0.26 eV is produced in this case.
Moreover, the mass gap can be opened in case of the absorbtion of
hydrogen on the graphene layer \cite{hydro} or due to finite size
effects (in nano-ribbons and so on).

In the present study we investigate the critical charge phenomenon
in gapped graphene and try to estimate the critical charge value.
The two-dimentional Coulomb problem without screening was studied
 in framework of Dirac equation formalism \cite{Khalilov} and
 in framework of the tight-binding lattice model \cite{lattice}. In
\cite{nero}, \cite{MIFI} the phenomenologically regularized
Coulomb problem was studied. In this paper we apply the effective
potential of the interaction between a charged defect and
 electron excitations, considering the $\sigma$-orbital screening
\cite{wehling1}, \cite{wehling2}.  It is well known that the
critical charge phenomenon is very sensitive to the details of the
potential near the position of the charge. So one can predict that
the value of the critical charge should be strongly dependent of
regularization of singular Coulomb potential near the origin. The
$\sigma$-orbital screening plays the role of the physical
regularization in this problem. We investigate the role of this
$\sigma$-orbital screening in this paper.

\section{The Dirac Equation of Gapped Graphene with Screened potential}

\begin{figure}[tbh!]
\centerline{\includegraphics[scale=0.5]{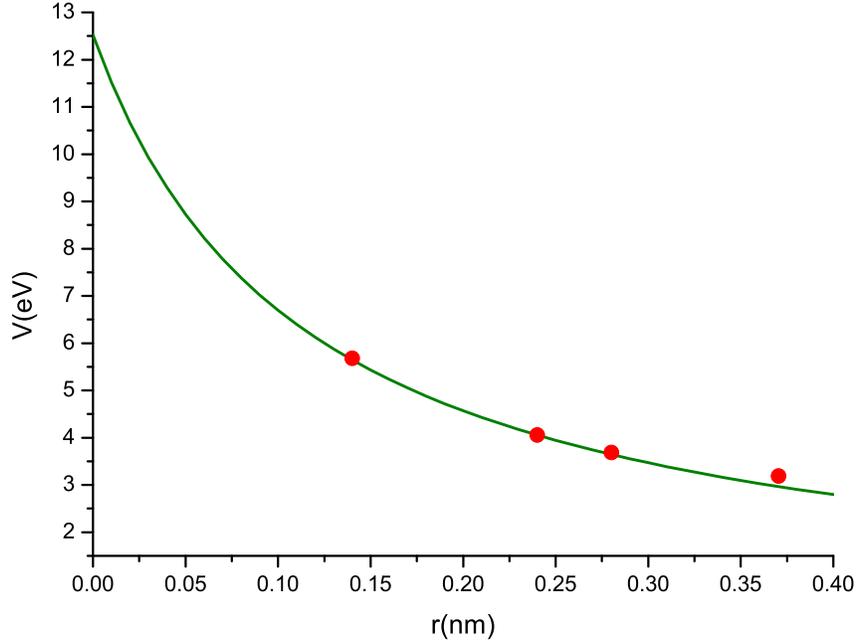}}
\caption{Screened potential for screening parameter $a=0.12$\,nm
(green line) and potential obtained by detailed calculation by
Wehling et al. (red dots). } \label{fig_r_V}
\end{figure}

Graphene is a two-dimensional material consisting of a hexagonal
(also called honeycomb) lattice of carbon atoms. Three of its four
valent electrons provide the chemical bonds in this lattice
($\sigma$-orbitals). The fourth one is valent and it provides the
electronic properties of graphene. The theory of solids predicted
for the excitations associated with electron tunnelling between
lattice sites to be massless Dirac particles. Later the mass of
the excitation turned up to depend on the substrate the graphene
sample placed on. For example,  Graphene on $\mbox{SiO}_2$
substrate is described by two-dimensional Dirac equation for
massless particles and graphene on SiC substrate is described by
Dirac equation for massive particles. The effective mass of these
particles is determined by energy gap $G=m v_f^2$. The value of
mass gap is determined by substrate, it is associated with
sublattice symmetry breaking. $v_f$ is Fermi velocity, an analogue
of speed of light for excitations in graphene. The value of Fermi
velocity is $v_f\approx c/300$. We represent the excitations as
two-dimensional Dirac particles with charge $e$ interacting with
three-dimensional Coulomb field of defect with charge $Z e$. The
Dirac equation for such particles is
\begin{equation}
\left(-i \hbar v_f  ( \sigma_x \partial_x +  \sigma_y
\partial_y ) - \frac{1}{\varepsilon_{eff}} \left( \frac{Z
e^2}{r+a}\right) + G{\sigma}_z \right)\psi(\mathbf{r}) = E \psi
(\mathbf{r}). \label{DiracEq}
\end{equation}
Here
\begin{equation}
\sigma_x =
\begin{pmatrix}
0\ &1\\
1\ &0
\end{pmatrix},\
\sigma_y =
\begin{pmatrix}
0&-i \\
i&0
\end{pmatrix},\
\sigma_z =
\begin{pmatrix}
1&0\\
0&-1
\end{pmatrix}
\label{sigma}
\end{equation}
are Pauli matrices. The effective dielectric permittivity is
$\varepsilon_{eff} = (\varepsilon + 1)/2$, where $\varepsilon$ is
the dielectric permittivity of the substrate. This expression is
motivated by the fact that the substrate is in only one half-space
from the graphene sheet, and there is vacuum ($\varepsilon=1$) in
the other half-space. The exact form of this expression is
strictly proved in classical electrodynamics with the method of
mirror charges \cite{Jackson}. ($\varepsilon=4$ for $\mbox{SiO}_2$
substrate and $\varepsilon=10$ for $\mbox{SiC}$ substrate). The
potential is screened by $\sigma$-orbitals. Fig. \ref{fig_r_V}
represents the screened Coulomb potential approximation for
$a=0.12$nm and values of the potential obtained from more detailed
condensed matter calculations in \cite{wehling1}, \cite{wehling2}.
One can see that the approximation is acceptable.

The critical charge problem in a similar (screened Coulomb)
one-dimensional potential has been studied theoretically
\cite{sveshnikov1}, \cite{sveshnikov2}, \cite{sveshnikov3}.

We explore the ground state near a charged defect. The experiments
on critical charge in graphene use charged ions placed on its
surface as such defects. These ions were discovered to form a
non-trivial distribution of electron-hole pairs. It confirms that
even a small charge is sufficient to be critical in this system.
We apply a quite simple theoretical model to describe the critical
charge in graphene placed on a substrate generating a mass gap.
This formulation of the problem resembles the one for critical
charge of atomic nuclei.

The equation (\ref{DiracEq}) can be solved with the substitution

\begin{equation}
\psi_j(r,\phi) = \frac{1}{ \sqrt{r} }
\begin{pmatrix}
e^{ -i \left(j - 1/2\right) \phi }A(r)
\\
ie^{ -i \left(j + 1/2\right) \phi }B(r)
\end{pmatrix},
\label{Wiev}
\end{equation}
where $j$ is the total angular momentum of the state.

We obtain a system of differential equations that provides the
solution for radial part

\begin{equation}
\begin{bmatrix}
\left( E + \frac{1}{\varepsilon_{eff}} (\frac{Z e^2 }{r+a}) - G
\right) & - \hbar v_f \left( \partial_r + \frac{j}{r} \right)
\\
\hbar v_f \left( \partial_r - \frac{j}{r} \right) & \left( E +
\frac{1}{\varepsilon_{eff}} (\frac{Z e^2 }{r+a}) + G \right)
\end{bmatrix}
\begin{bmatrix}
A(r)
\\
B(r)
\end{bmatrix}
= 0. \label{MySystem}
\end{equation}



We consider the ground state (s-wave) electron (j=1/2). The Dirac
equation for this case is solved by shooting method. The
performance, accuracy and even correctness of such algorithms
strongly depend both on the physical parameters of the system and
on the parameters of the algorithm. In order to control all of
these effects we used an own written program for numerical
calculations.

Idea of the shooting method is numerical integration of
differential equation from zero and from infinity and stitching
the solutions at some middle point. Asymptotic behavior of the
correct solutions is $A(r), B(r)\sim \sqrt{r}$ at $r\to 0$ and
$A(r), B(r)\sim e^{-\kappa r}$ at $r\to\infty$. Corresponding
boundary conditions in zero are set by hand. To obtain the
required asymptotic at infinity it is enough to start shooting
from arbitrary (small) value of function at arbitrary (large)
value of radius, than the correct solution exponentially increases
and the incorrect one exponentially deceases.

The integration of the equation was performed by RK4 method. The
criterion of correct (smooth) stitching is
$W(E)=A_lA'_r-A_rA'_l\to 0$ (the similar expressions as for $A(r)$
are suggested everywhere for $B(r)$). Here indices $l$ and $r$
stand for left (shooting from zero) and right (shooting from
infinity) solutions in the stitching point for given energy. This
equation was solved with the accuracy $10^{-6}$ for $E$.

To control that the stitching is really smooth we also used a
criterion $(A_lA'_r-A_rA'_l)/(A_lA'_r+A_rA'_l)\ll10^{-3}$.

The numerical solution is closer to the physical one for smaller
discretization step $d$ and larger "infinity" $R_\infty$ (the
maximum value of radius, a point where the boundary conditions for
infinity are set). We suggested that the numerical solution is
sufficiently close to the continuous one if 10 times smaller $d$
and 10 times larger $R_\infty$ does not significantly change it.

Another criterion of the correctness of the solution is based on
the exponential decay of the bound state wavefunction: we
suggested that the numerical solution is correct if
$\overline{|A(r)|}_{[0.9R_{\infty};R_{\infty}]}/ \mathrm{max}
A(r)_{[0;R_{\infty}]}\ll10^{-3}$.

The solution provides the ground
state wavefunction and energy level for the equation
(\ref{MySystem}). We examine the energy level inside the mass gap
$-G < E < G $ as a function of the charge of the defect and the
dielectric permittivity of substrates $\varepsilon$.

\section{Critical charge in gapped graphene for screened potential model}

The equation (\ref{MySystem}) was solved as an eigenvalue problem
to obtain  the spectrum energy levels $E_i$ and corresponding
wavefunction. In fact we need only the ground state energy level
$E_0$ to determine the critical charge. The spectrum was obtained
for various values of defect charge $Z e$ and the dielectric
permittivity of substrates $\varepsilon$, the calculations were
performed for a value of parameters $a = 0.12$ nm (screening
length in potential) and $G=m v_f^2$ (mass gap, i.e. excitation
effective mass).


At large value of the dielectric permittivity  of the substrate
the potential of interaction between the defect and the electron
excitation is deceased by the influence of the substrate and all
the energy levels, including the ground state one, asymptotically
go to $+G$. The energy levels decrease if the dielectric
permittivity  of the substrate becomes smaller. The decrease of
ground state energy to $-G$ corresponds to the critical charge
$Z_{cr}$ -- at these values of defect charge and the dielectric
permittivity  of the substrate the energy production of the
electron-hole pair is equals to zero and our system becomes
unstable. Physically it leads to production of the the
electron-hole pairs and to practically screening of the
critical charge of defect by the cloud of inverse charges. This
process could lead to the generation of "additional" electrons and
to changing of the local density of states which can be detected
experientially and can be interpreted as a signal of the critical
charge achievement.


In our work the two type of experimental situation are studied.
Let us consider the defects with different integer value of charge
(Z=1, 2, 3, 4, 5) placed on the graphene sheet. For the generation
of mass gap the special type of the substrate must be used. As we
discussed in the previous section there are two types of substrate
that can break the sublattice symmetry of graphene and generate
the mass gap in this system. One of them is the single layer boron
nitride ($\mbox{BN}$) on the ($\mbox{SiO}_2$) substrate and
another is barrier graphene layer on silicon carbide
($\mbox{SiC}$) substrate. Let us consider each of these
situations.

The experimental scheme for the $\mbox{BN}$-$\mbox{SiO}_2$ system
is shown in the Fig. 2. The dielectric permittivity of
$\mbox{SiO}_2$ is 3.9. The difference of boron-carbon
nitrogen-carbon interactions generates the mass gap for electron
excitations in graphene about $G=0.053eV$. The results of
numerical simulation of this system for different values of the
defect charge and of the substrate dielectric permittivity are
shown in the Fig. 2.

\begin{figure}[tbh!]
\centerline{\includegraphics[scale=0.5]{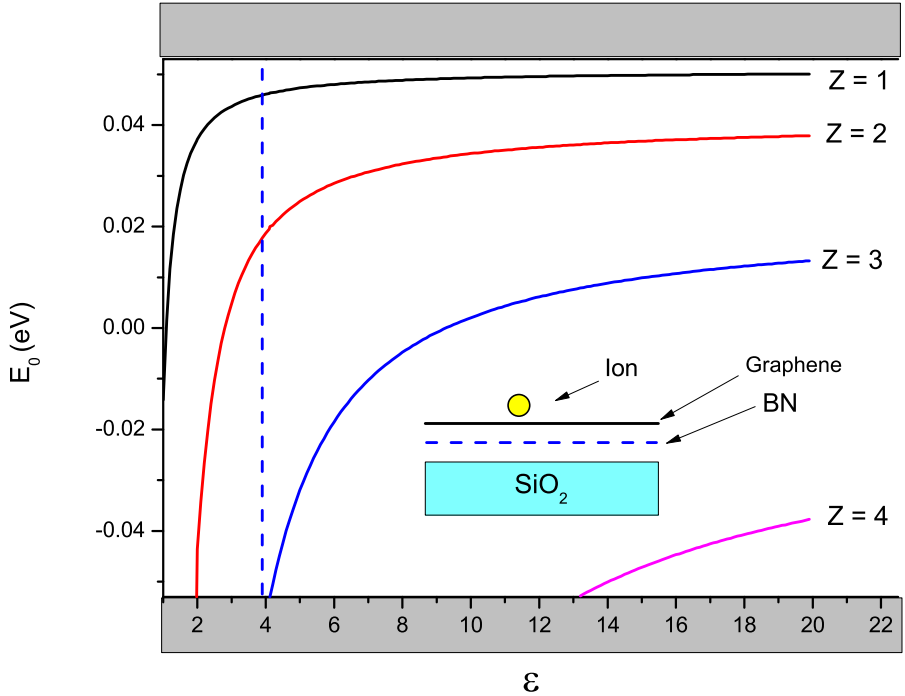}}
\caption{Ground state energy as a function of the substrate
dielectric permittivity $\varepsilon$ for different charges of the
defect. Mass gap is $G=0.053$ eV (boron nitride). The dashed line
corresponds to the dielectric permittivity of $\mbox{SiO}_2$.}
\label{fig_g_E0_a}
\end{figure}

One can see that for $\mbox{BN}$-$\mbox{SiO}_2$ system the charges
3 and more are super-critical, and the charge 3 is very close to
the critical value.

The experimental scheme for the barrier graphene-SiC system is
shown in the Fig. 3. The dielectric permittivity of $\mbox{SiC}$
is about 10. As the upper graphene layer is in AB-structure with
the barrier graphene layer, one of the sublattices of the layers
is in strong interaction while the other is not. This is a strong
sublattice symmetry breaking that generates a mass gap of about
$G=0.26$ eV. The results of numerical simulation of this system
for different values of the defect charge and of the substrate
dielectric permittivity are shown in the Fig. 3.

\begin{figure}[tbh!]
\centerline{\includegraphics[scale=0.5]{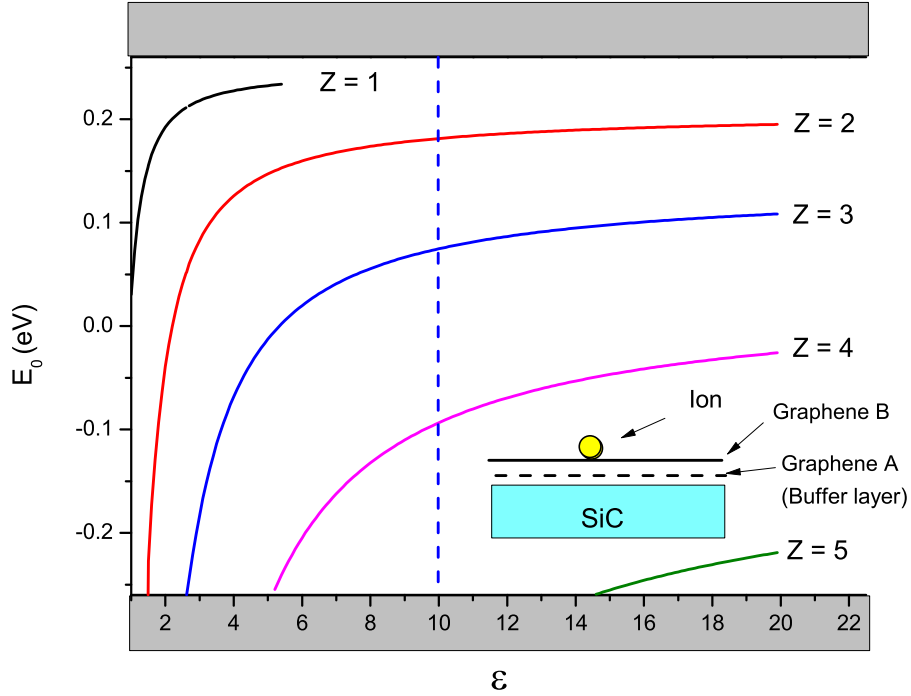}}
\caption{Ground state energy as a function of the substrate
dielectric permittivity $\varepsilon$ for different charges of the
defect. Mass gap is $G=0.26$ eV (barrier graphene layer). The
dashed line corresponds to the dielectric permittivity of
$\mbox{SiC}$.} \label{fig_g_E0_a}
\end{figure}

The simulation shows that for barrier graphene-SiC system the
charges 5 and more are super-critical; and the charge 4 and less
are sub-critical.

\section{Conclusions}

The main goal of our work is to study the critical charge effect
in the gapped graphene. We have investigated the influence of the
$\sigma$-orbitals screening on this phenomenon. It is well known
that the critical charge value is very sensitive to the behavior
of the interaction potential near the charge. Due to this fact one
could expect that the parameters of screening are crucial for this
problem.

The recent experiment  \cite{Nature} has shown that the external
charge vacancy in graphene can form the artificial atom with
electrons of conductivity. It was shown that the transition to the
supercritical regime can be observed in this experiment by the
study of the density of state peaks below the Dirac point energy.
These peaks are associated with quasi-bound states localized near
the charge vacancies so they are the clear signal of the
transition into the supercritical regime.

It is very well known that the supercritical charge effect is very
sensitive to the behavior of the potentials at a short distance to
the charge. The great success in the experimental study of the
supercharge phenomenon in graphene opens the possibility of the
direct investigation of short distance effects in graphene such as
the Coulomb potential screening, the influence of the dielectric
properties of the substrate, the effects of graphene sheet
deformation, and so on. In our study we have investigated the
influence of the $\sigma$-orbital screening of Coulomb potential
on the critical charge effect in the cases of two very popular
gapped graphene systems: graphene on the
$\mbox{BN}$-$\mbox{SiO}_2$ substrate and graphene on barrier
graphene-SiC substrate.

In our work we used the 2+1-dimensional Dirac equation for the
study of the excitation in graphene in strong external potential
of a point charge. Based on the \cite{wehling1}, \cite{wehling2}
the potential was suggested to be Coulomb screened. This screening
was obtained in the framework of the effective extended Hubbard
model of graphene by using of determinant quantum Monte Carlo
(DQMC) method. The screening of Coulomb potential was also studied
in \cite{TF} in framework of the Thomas-Fermi method. This
approximation method is very well applicable for a wide range of
radius values, and the resulting form of screening differs from
DQMC prediction we used in our work.  We plan to investigate a
similar problem also for  this form of potential in our future
studies. The comparison of the results with the experimental data
can provide the  most optimal form of the approximation of Coulomb
potential at short distances.

 Dirac equation was solved numerically by the shooting method. We
have investigated two physical situations: graphene on the
$\mbox{BN}$-$\mbox{SiO}_2$ substrate and graphene on barrier
graphene-SiC substrate. The numerical simulations have shown that
the values of critical charge are 3 and 5 respectively.

We also investigate how the substrate dielectric permittivity does
influence on the critical charge phenomenon. For any physical
($\varepsilon>1$) situation the critical charge is larger than 1.
This conclusion is in agreement with the experimental data
\cite{Nature}.

\section{Acknowledgements}

 This work has been supported by  the grant from the Russian Science
Foundation (project number 16-12-10059).


\newpage







\end {document}